\documentclass[a4paper,11pt]{article}
\usepackage{jcappub} 
\usepackage{lineno}
\usepackage{orcidlink}

\newcommand{\dneff}{\Delta N_{\text{eff}}}
\newcommand{\neff}{N_{\text{eff}}}
\newcommand{\SM}{\text{SM}}
\newcommand{\abs}[1]{\left| #1 \right|} 
\newcommand{\zprime}{{Z^{\prime}}}
\newcommand{\dps}[1]{\frac{d^3 p_{#1}}{(2\pi)^3 2 E_{#1}}}

\title{\boldmath\huge Improved cosmological limits on  $Z^\prime$ models with light right-handed neutrinos}

\author[a]{Tim Herbermann\,\orcidlink{0009-0003-4771-5759},}
\author[a]{Manfred Lindner\,\orcidlink{0000-0002-3704-6016}}
\affiliation[a]{Max-Planck-Institut für Kernphysik,\\
Saupfercheckweg 1, 69117 Heidelberg, Germany}

\emailAdd{tim.herbermann@mpi-hd.mpg.de}
\emailAdd{lindner@mpi-hd.mpg.de}

\abstract{
We improve limits on $\zprime$ extensions of the Standard Model (SM) with light right-handed neutrinos. The presence of shared gauge interactions between the light right-handed neutrinos and other SM fermions allows for production of $\nu_R$ in the early Universe and we use the excess in the effective number of neutrino species $\Delta N_\text{eff}$ to place limits. Our benchmark model is a minimal gauged $U(1)_{B-L}$ that often arises as a building block in other models, and we discuss applicability to more general $U(1)$ extensions. 
We devise an improved Monte Carlo integration scheme convenient for implementation of generic integrated Boltzmann equations with minimal simplifying assumptions. We sketch our numerical implementation in detail for future reference.
Using the new ACT DR6 limit $\Delta N_\text{eff}<0.17$, we improve constraints on the gauge coupling for $1\mathrm{\,GeV} < m_{Z^\prime} < 100\mathrm{\,TeV}$ by orders of magnitude and find the strongest limits thus far, surpassing even current and future colliders, and explore the potential of future CMB experiments to test $U(1)$ extensions up to the GUT scale.
We perform a detailed analysis of the robustness of cosmological limits within standard and non-standard thermal histories and find that a strong first order phase transition, early dark energy or early matter domination could dilute $\nu_R$ abundances beyond detection.
We investigate the effect of reheating on $\nu_R$-genesis and provide results and prescriptions to apply our bounds to non-standard thermal histories. Limits are generically weakened for reheating $T_\text{reh}\ll m_{Z^\prime}$. Our results suggest that projected limits on $Z^\prime$ with Dirac neutrinos can only be accommodated for in non-standard thermal histories, thus limiting the options to include dark matter candidates or Dirac leptogenesis. 
}

\begin{document}
\maketitle
\flushbottom

\section{Introduction}

The nature of neutrinos is one of the big open questions in neutrino physics. While the Standard Model (SM) of particle physics predicts massless neutrinos, neutrino oscillations and the implied non-vanishing neutrino masses are an experimental reality (see e.g. \cite{Giganti:2017fhf}). What these observations do not yet reveal is whether the active neutrinos are of Majorana or Dirac type. A possible observation of neutrinoless double beta decay \cite{Dolinski:2019nrj} will clarify the situation -- with profound consequences beyond neutrino mass model building. As far as cosmology is concerned, one major difference between Dirac and Majorana neutrinos is that the former come with light right-handed neutrinos that could give an extra contribution to the total radiation budget of the Universe.

Cosmological excess radiation is conventionally normalized to the energy density of one active neutrino flavor, $\neff = (8/7)\,(11/4)^{4/3} \rho_\text{rad}/\rho_\gamma$. The theoretical SM expectation is $\neff = 3.044$ \cite{Froustey:2020mcq,EscuderoAbenza:2020cmq,deSalas:2016ztq,Akita:2020szl}\footnote{Current theoretical predictions prefer this value, with some concurrent studies preferring $\neff = 3.045$. We adopt the aforementioned value, but note that this small difference will not affect our conclusions in a meaningful way.}, and deviations from this expectation are usually parametrized as $\dneff=  \neff-3.044$. This is a powerful probe of any SM extension that introduces new light degrees of freedom.

If the extra degrees are in chemical equilibrium in the early Universe, their relic abundance and contribution to the effective number of neutrinos can readily estimated by their decoupling temperatures as \cite{Abazajian:2019oqj}
\begin{equation}
    \dneff  \simeq 0.027\,g_x \left(\frac{106.75}{g_\star(T_\text{dec})}\right)^{4/3}\,,
    \label{eq:basis}
\end{equation}
where $g_x$ is the (effective) internal degrees of freedom for a light species $x$. The presence of three $\nu_R$ that left equilibrium with the SM at the electroweak scale or above would contribute a significant $\dneff\simeq 0.14$. However, it is possible that interactions are too feeble to ever thermalize light right-handed neutrinos, and akin to the often studied freeze-in production of dark matter \cite{Hall:2009bx}, a significant relic abundance can still be realized \cite{Luo:2020fdt,Adshead:2022ovo}.

Bounds on  Dirac neutrinos in context of $Z^\prime$ extensions more generally and gauged $U(1)_{B-L}$ in particular have been studied before, with limits based on the Planck best fits \cite{Planck:2018vyg} by adopting either simplified scattering rate approaches or more sophisticated calculations involving Boltzmann equations. The considered models constitute both minimal extensions \cite{Heeck:2014zfa,Abazajian:2019oqj,Adshead:2022ovo,Esseili:2023ldf,Caloni:2024olo} and extended models that may include dark matter candidates, or address other problems \cite{Biswas:2022fga,Babu:2022ikf,Borah:2022obi,Berbig:2022pye,Mishra:2021ilq,Borah:2025fkd,Berbig:2022nre,Mahanta:2021plx,Han:2020oet,Das:2023yhv,Adshead:2020ekg}. It is possible that extended sectors also affect $\dneff$ in non-trivial ways, even reducing it in some circumstances \cite{Safi:2024bta,Wang:2023csv}. Such studies typically assume a standard thermal history, a plausible but strong assumption that deserves scrutiny and that was already partially relaxed in \cite{Caloni:2024olo}.

In what follows, we present an improved calculation of the relic abundance of right-handed neutrinos under the benchmark assumption that $B-L$ is a gauged symmetry. The particle model we consider is minimal, but arises as a building block in all models that have Dirac neutrinos and gauged $B-L$\footnote{While we only consider a Dirac nature of neutrinos here, our results also apply to light right-handed neutrinos in the Majorana case as long as $m_{\nu_R} < T_\text{CMB}$ and the $\nu_R$ contribute as cosmological radiation.}. The $\nu_R$ abundance that arises from this building block can thus be considered an irreducible contribution that deserves extra scrutiny. Similar structures will naturally arise in all models that introduce Dirac neutrinos that share gauge interactions with at least a subset of the SM. Therefore, results we find within our benchmark study will approximately hold also for any other $Z^\prime$ extension that couples $\nu_R$ and SM fermions.

We calculate the excess radiation from $\nu_R$ by solving the respective Boltzmann equations. Additional focus is given to the cosmological assumptions and we analyze in detail the effects of non-standard thermal histories. The framework we develop is exact on the level of integrated Boltzmann equations with only minimal additional assumptions. Key part of our computation is the efficient Monte Carlo integration of the relevant collision operators, and we incorporate in particular leading thermal corrections and the effects of the final state statistics from the finite density environment. While it is not expected that these effects give rise to significant corrections for $\nu_R$-genesis, they are straightforward to include in Monte Carlo approaches. The ansatz we sketch should prove useful for the study of other systems of Boltzmann equations beyond production of $\nu_R$, where such effects can be significant.  

In Section \ref{sec:dneff}, we summarize the current experimental state of $\dneff$, and how it relates to Dirac neutrino model building. We introduce the minimal model for this study in Section \ref{sec:model}. Then, we develop an almost approximation-free approach to integrated Boltzmann equations in Section \ref{sec:formalism}, drawing on the ansatz in \cite{Luo:2020fdt} and developing it further. We discuss our improved limits and forecasts in Section \ref{sec:results}, including a comprehensive discussion of their robustness and required modifications beyond the standard cosmological picture. We conclude in Section \ref{sec:conclusions}.

\section{Dirac neutrinos in the $\dneff$ landscape}
\label{sec:dneff}

If neutrinos are Dirac particles, the light right-handed extra degrees of freedom contribute to the radiation budget of the Universe. Excess radiation is constrained by precise measurements of the angular scale of acoustic peaks in the cosmic microwave background (CMB) and its damping tail \cite{2013PhRvD..87h3008H}, and by the primordial abundance of elements from Big Bang Nucleosynthesis (BBN).
Proposals to use either CMB features or primordial element abundances to constrain extra radiation are old and have been considered e.g. in \cite{PhysRevLett.43.239,Dolgov:2002wy,RevModPhys.53.1}. Since then, the bounds on the effective number of neutrino species $\dneff$ tightened significantly -- with Planck indicating $\dneff < 0.285$ at $95\%$ C.L. \cite{Planck:2018vyg,Abazajian:2019oqj}, cosmology provided valuable input for models that predict new light degrees of freedom. 

Recently, the data release 6 of the Atacama Cosmology Telescope (ACT)\cite{ACT:2025tim,ACT:2025fju} has improved on these limits. Utilizing the full primary CMB data from ACT in combination with large and medium scale data from Planck, the limit is pushed down to $\dneff<0.17$ at $95\%$ C.L., the strongest limit on relativistic relics so far.

These results are complimented by joint CMB+BBN analyses, which combine data sets across the cosmological history and find agreeable limits of $\dneff < 0.180$ \cite{Yeh:2022heq} and $\dneff<0.163$ \cite{Fields:2019pfx} at $95\%$C.L. respectively. An incomplete subset combination of SPT-3G polarization and lensing data, when combined with Planck, ACT and BAO, finds $N_\text{eff}=2.86\pm0.13$, $N_\text{eff}=2.83\pm0.13$, and $N_\text{eff}=2.89\pm0.23$ when allowing for extended $\Lambda$CDM$+N_\text{eff}$, $\Lambda$CDM$+N_\text{eff}+\sum m_\nu$, and $\Lambda$CDM+$N_\text{eff}+Y_p$ fits respectively \cite{SPT-3G:2024atg}, giving an indication that cosmological data continues to disfavor large amounts of extra radiation and future searches will continue to tighten constraints on SM extensions with new light degrees of freedom.

Forecasted sensitivities of ongoing and planned experiments surpass these constraints, and Simons Observatory (SO) \cite{2019BAAS...51g.147L} and SPT-3G \cite{SPT-3G:2014dbx} are projected to reach around $\dneff < 0.12$ at $95\%$C.L. \cite{Abazajian:2019oqj}, and CMB-S4 \cite{Abazajian:2019eic} anticipates to reach $\dneff<0.06$. Futuristic proposals like CMB-HD \cite{Sehgal:2019ewc} are designed with the goal of breaching the important $\dneff = 0.027$ benchmark of a single thermalized Nambu-Goldstone boson. We conclude that current cosmological bounds are close to the important benchmark of three fully thermalized $\nu_R$, and with the indication that ongoing or upcoming next-generation experiments will push below this important threshold soon.

Assuming that the Dirac mass is generated by the SM Higgs, and that the interactions of the right-handed partners are restricted to Yukawa interactions, current neutrino mass bounds imply $\dneff < 10^{-11}$ \cite{Luo:2020fdt}, a contribution too small to ever be within experimental reach. However, it is generally expected that the neutrino sector encompasses larger extensions. First, to explain the smallness of neutrino masses when compared to the SM fermions, and second, to preserve their Dirac nature if applicable. The method of choice is usually by means of a (global or local) symmetry that forbids the presence of a Majorana mass term. While it is perfectly fine to protect the Dirac nature by means of a global symmetry, the possibility of quantum gravity to break global symmetries would suggest that a protection by means of a gauge symmetry is the preferred option. Among the most popular and well motivated options is the promotion of the accidental $U(1)_{B-L}$ symmetry of the SM to a local one. This unavoidably necessitates the introduction of $\nu_R$ to cancel gauge anomalies, thus making the presence of right-handed neutrinos a requirement for internal consistency.
We note here that $B-L$ extensions of the SM, although well motivated, are already subject to strong laboratory constraints even beyond the Dirac hypothesis (e.g. \cite{Escudero:2018fwn}).

\begin{figure}
    \centering
    \includegraphics[width=1.\linewidth]{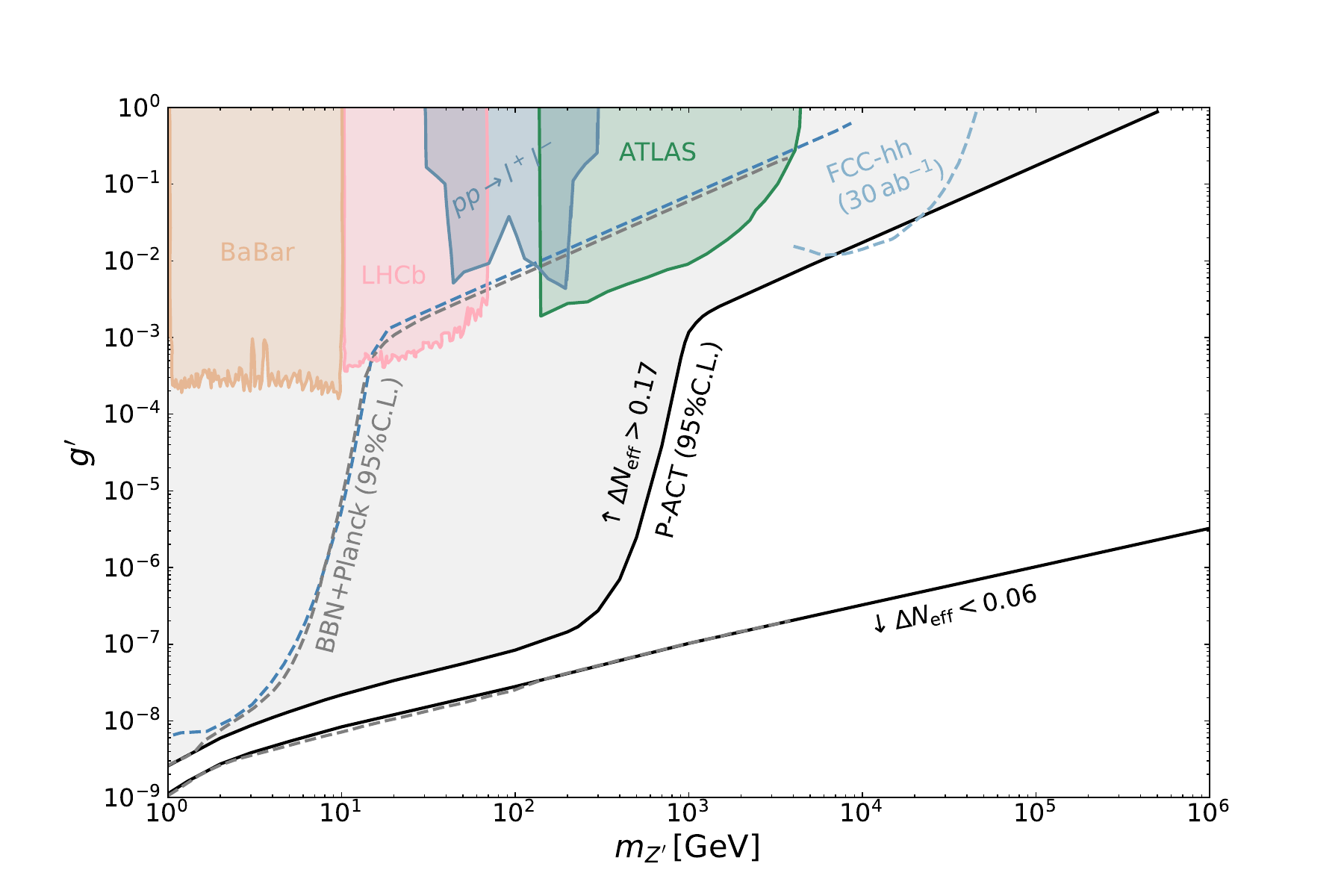}
    \caption{Constraints on $U(1)_{B-L}$ gauge bosons in presence of light right-handed neutrinos. The black solid curves are updated contours of $\dneff$. The upper one corresponds to $\dneff=0.17$, excluded by ACT \cite{ACT:2025fju} and disfavored by current BBN+CMB combinations \cite{Yeh:2022heq,Fields:2019pfx}. The lower curve is $\dneff=0.06$, which serves as a benchmark for future CMB experiments (see Section \ref{sec:dneff}). We also show previous forecasts and Planck+BBN limits from References \cite{Adshead:2022ovo}(grey dashed) and \cite{Abazajian:2019oqj}(blue dashed). We show complimentary limits from ATLAS \cite{ATLAS:2017fih}, limits from the Drell-Yan process in ATLAS \cite{ATLAS:2017rue,Escudero:2018fwn}, a recasting of LHCb limits \cite{LHCb:2017trq,Escudero:2018fwn}, and dark photon decays from BaBar \cite{BaBar:2014zli,Escudero:2018fwn}. For comparison, we show the projected reach of FCC-hh for $Z^\prime$ searches from \cite{Liu:2022kid}.}
    \label{fig:limits}
\end{figure}

\section{Minimal gauged $B-L$ with Dirac neutrinos}
\label{sec:model}

Our benchmark model for $Z^\prime$ extensions is a gauged $U(1)_{B-L}$, and we adopt a minimal setup. The derived constraints will either be exact in the case that additional model constituents are feebly coupled, or conservative, as extra interactions tend to further boost production of $\nu_R$. Therefore, we introduce three right-handed partners $\nu_R$ to the SM neutrino and augment the Lagrangian density by
\begin{equation}
    \mathcal{L}\supset \frac{1}{4}F^\prime_{\mu\nu}F^{\prime\mu\nu} + \frac{1}{2}m_\zprime^2 Z^{\prime}_\mu Z^{\prime\mu} + g^\prime  Z^\prime_\mu \sum_{f\in\mathrm{SM}} Q^{(f)}_\mathrm{B-L}  \bar f \gamma^\mu  f + g^\prime  Z^\prime_\mu Q^{(\nu_R)}_\mathrm{B-L}  \bar \nu_R \gamma^\mu  \nu_R\,,
    \label{eq:lagrangian}
\end{equation}
and made explicit the interactions with SM fermions (including the left neutrino) and the light right-handed neutrinos. The SM $B-L$ gauges are $1/3$ for quarks and $-1$ for leptons. In the absence of other fermions charged under $U(1)_{B-L}$, we are restricted to choose either $Q^{(\nu_R)}_{B-L} = -1$ or  $Q^{(\nu_R)}_{B-L} = -4,-4,5$ to cancel gauge anomalies. We remain agnostic about the origin of the Dirac mass for neutrinos. For the $-1$ choice of gauge charges, it is possible to write down Yukawa terms with the Higgs doublet. The Dirac nature of the neutrino is then protected by the gauge symmetry and the model as it stands can be considered complete up to the origin of the $\zprime$ mass. Whether it is realized as a Stueckelberg mass, or if an extended scalar sector breaks $\Delta(B-L)\neq 2$ and induces a mass term for the $\zprime$ is largely irrelevant to obtain robust constraints on this class of models.
In the case of $Q^{(\nu_R)}_{B-L} = -4,-4,5$, a Dirac mass from the Higgs only is not possible. This necessarily implies that such a model is incomplete already on the level of $\nu_R$ mass generation. Still, it would be possible to derive constraints that are conservative, or exact if the extended sector is largely decoupled from the SM and $\nu_R$ production. We will not explicitly compute limits for this particular case, but approximate limits can be obtained from appropriate rescaling of the gauge coupling. Likewise, other $\zprime$ extensions can also be approximately constrained by rescaling.

If $B-L$ remains unbroken, the Lagrangian in Equation \ref{eq:lagrangian} should contain kinetic mixing terms $\mathcal{L} \supset \frac{\epsilon}{2} Y_{\mu\nu}  F^{\prime\mu\nu}$, since such a mixing term will inevitably generated from radiative corrections. As far as mixing to the photon is concerned, exact cancellations from destructive interference are fine-tuned and not stable under running, and cancellations from mixing to the $Z$ cannot be exact due to the chiral nature of the latter. Thus, if $B-L$ remains unbroken and kinetic mixing is necessarily generated, it can at most provide a minor correction of the effective coupling that we probe, and the effect can again mostly be absorbed in a rescaling of gauge coupling.

\section{Boltzmann equations}
\label{sec:formalism}
The abundance of cosmological relics is conveniently calculated in the framework of Boltzmann equations (e.g. \cite{Kolb:1990vq}). For our purposes, it is not necessary to study Boltzmann equations on phase space level, as results of integrated Boltzmann equations for number densities or energy densities usually give excellent results compared to phase space solutions (see e.g. \cite{Ala-Mattinen:2022nuj}).
In the following, we sketch an alternative numerical approach to integrated Boltzmann equations without further approximations by means of Monte Carlo integration of the collision operator. This is complementary to available numerical schemes (e.g. \cite{Belanger:2018ccd}) that also aim to include relativistic effects and quantum statistics. While such schemes are well suited for computational efficiency, we can allow to make a few drawbacks on performance but enjoy a more straightforward implementation. We draw in particular on Reference \cite{Luo:2020fdt}, and develop the Monte Carlo approach further. More details on the numerical implementation are given in Appendix \ref{sec:integrations}.

\subsection{General framework}
We simplify the problem by treating the right-handed neutrinos and anti-neutrinos as a common species with $g_{\nu_R}=2\times 3$ internal degrees of freedom \cite{Adshead:2022ovo,Luo:2020fdt,Luo:2020sho}. As long as $\zprime$ couples (approximately) universal to $\nu_R$, this is a good approximation.
The Boltzmann equation for the energy density of the collective $\nu_R$ species and the SM are given by
\begin{equation}
    \begin{aligned}
    \frac{d\rho_{\nu_R}}{dt} &= -4H\rho_{\nu_R} + \mathcal{C}^{(\rho)}_{\SM\rightarrow\nu_R}\,,\\
    \frac{d\rho_\SM}{dt} &= -3H(\rho_\SM+P_\SM) - \mathcal{C}^{(\rho)}_{\SM \rightarrow \nu_R}\,.
    \label{eq:energy_density}
    \end{aligned}
\end{equation}
Here, the integrated collision operator describes the energy transfer rate from the SM plasma to the $\nu_R$ sector. The Hubble rate is $H^2=\frac{8\pi G}{3} (\rho_\SM + \rho_{\nu_R})$. In its most general form, the collision operator can be written as a sum over any number of initial states $i$ and final states $f$ that contain at least one $\nu_R(\bar\nu_R)$
\begin{equation}
\begin{aligned}
    \mathcal{C}^{(\rho)}_{i\rightarrow f} =& \sum_{i,f \ni \nu_R} \int d\Pi_{i_1}...d\Pi_{i_n} d\Pi_{f_1}...d\Pi_{f_m}
    (2\pi)^4 \delta^{(4)}\left(\sum p_i -\sum p_f\right) E_{\nu_R} \\
    \times & \left( \abs{\mathcal{M}_{i\rightarrow f}}^2 f_{i_1}...f_{i_n}\bar f_{f_1}...\bar f_{f_m}  - \abs{\mathcal{M}_{f\rightarrow i}}^2  f_{f_1}...f_{f_m}\bar f_{i_1}...\bar f_{i_n} \right)\,.
    \label{eq:coll_rho}
\end{aligned}
\end{equation}
Here we introduced several shorthands, that we will use frequently. We denote the Lorentz invariant phase space element as $d\Pi_i = \frac{d^3p_i}{(2\pi)^3 2E_i}$, the phase space density of $x$ is $f_x=f_x(E_x)$, and $\bar f_x = 1\pm f_x$ is the final state phase space and the sign depends on whether particle $x$ is a boson (fermion). We will further assume $CP$ symmetry such that $\abs{\mathcal{M}_{i\rightarrow f}}^2 = \abs{\mathcal{M}_{f\rightarrow i}}^2$ holds. Note that in our convention, the matrix element already contains the symmetry factor $1/n!$ for identical particles in the initial and final states \cite{Kolb:1990vq}. We restrict the analysis to $2\leftrightarrow2$ and $2\leftrightarrow 1$ processes, but a generalization to $n\rightarrow m$ processes would be straightforward. Further, we assume that the SM remains in thermal equilibrium at all times\footnote{It would be interesting to consider cases where e.g. resonantly enhanced processes could cause brief violations of the equilibrium assumption. This would need to be studied on the phase space level and is beyond the scope of this work.}.

If the SM maintains equilibrium, then
\begin{equation}
    \rho_\SM=\frac{\pi^2}{30}g_{(\rho)}T^4\,,\quad P_\SM=\frac{\pi^2}{90}g_{(P)}T^4\,,
\end{equation}
where we allow for the small difference between $g_{(\rho)}$ and $g_{(P)}$ that occurs during phase transitions or if a constituent becomes non-relativistic and disappears from the plasma \cite{Husdal:2016haj}, and we use the numerical values from Reference \cite{Laine:2015kra}.

To close the system of equations, we prescribe equilibrium phase space distributions. By assumption, we can incorporate equilibrium distributions for all SM particles. We make the same prescription for $\nu_R$, and we compute the temperature from inverting the energy density of $\nu_R$ under the equilibrium assumption \cite{Luo:2020fdt,Adshead:2022ovo}. For as long as the right-handed neutrinos are chemically coupled to the SM or maintain internal equilibrium, this approach is exact. During the decoupling process from the SM plasma, the distribution maintains its equilibrium shape up to spectral distortions. In the opposite limit, where $\nu_R$ never attains chemical equilibrium with the SM, the backreaction and final state occupation is suppressed and insensitive to the underlying phase space density, thus making this choice a reasonable approximation in both limits. Only an intermediate regime, close to thermalization of $\nu_R$ is sensitive to the precise phase space distribution, but our prescription can be understood as a smooth interpolation between the two limiting cases.

We integrate the collision operator numerically without any additional approximations. This makes the result exact on the level of integrated Boltzmann equations and the phase space density prescription we adopt. This can be done by performing a subset of the integrals in the collision operator analytically, and then evaluating the remaining integrals numerically. Since the integrals are still highly dimensional, it is advantageous to resort to Monte Carlo integration \cite{Luo:2020fdt}. The details of the numerical implementation are given in Appendix \ref{sec:integrations}, and we find an acceptable performance given the ease of implementation. This complements previous approaches to production of light right-handed neutrinos in the early Universe, where the quantum statistics of final states in particular are often neglected and simplified prescriptions to include the backreaction are employed \cite{Adshead:2022ovo,Luo:2020fdt,Arcadi:2019oxh,DeRomeri:2020wng,Lebedev:2019ton}, or complete but rather complicated schemes to include also the quantum statistics of final states \cite{Belanger:2018ccd,Bringmann:2021sth}. We stress again that with straightforward numerical integration we can consistently include final state statistics and the backreaction as it appears in the collision operator. However, in the scenario we study here we do not expect these corrections to be significant.

We integrate the set of Boltzmann equations Equation \ref{eq:energy_density} from an initial reheating temperature $T_\text{reh}$ up until the energy transfer is completed ($\mathcal{C}/H\ll 1$). Then, $\dneff$ can be inferred from the relation \cite{Luo:2020fdt}
\begin{equation}
    \dneff \simeq \frac{4}{7} g_{(\rho)} \left(\frac{10.75}{g_{(s)}}\right)^{4/3} \frac{\rho_{\nu_R}}{\rho_\SM}\,,
\end{equation}
which holds for all temperatures after the $\nu_R$ abundance is set and $\nu_R$ and the SM are no longer in contact.

\subsection{Finite temperature effects}
In a dense plasma, the dispersion relation of particles is modified by finite temperature corrections. While a full treatment at the level of thermal quantum field theory is beyond the scope of our approach here, the leading effect is often captured by introducing thermal masses \cite{DeRomeri:2020wng,Bringmann:2021sth,Elmfors:1993re}. We adopt this prescription here to account for leading thermal effects. To this end, all masses are replaced by the corresponding thermal masses in a plasma. Before electroweak symmetry breaking (EWSB), the thermal masses for the fermions are chiral, giving different corrections to left- and right-handed fields \cite{Elmfors:1993re} 
\begin{align}
    m_{l,L}^2 &= \frac{m_Z^2+2m_W^2+m_l^2}{2v_H^2} T^2\,,\\
    m_{l,R}^2 &= \frac{m_Z^2-m_W^2+\frac{1}{2}m_l^2}{2v_H^2} T^2\,,\\
    m_{q,L}^2 &= \frac{1}{6}g_s^2T^2 + \frac{3m_W^2+\frac{1}{9}(m_Z^2-m_W^2)+m_u^2+m_d^2}{2v_H^2}\,,\\
    m_{u,R}^2 &= \frac{1}{6}g_s^2T^2+\frac{\frac{4}{9}(m_Z^2-m_W^2)+\frac{1}{2}m_u^2}{2v_H^2}\,,\\
    m_{d,R}^2 &= \frac{1}{6}g_s^2T^2+\frac{\frac{1}{9}(m_Z^2-m_W^2)+\frac{1}{2}m_d^2}{2v_H^2}\,,
\end{align}
and after EWSB, the Higgs induced mass dominates but is corrected by \cite{Bringmann:2021sth}
\begin{align}
    \Delta m_l^2 &=\frac{1}{8}e^2T^2\,,\\
    \Delta m_q^2 &= \frac{1}{6}g_s^2T^2\,.
\end{align}
Note that in principle we would have to add the contributions from the new $\zprime$ boson, whenever it is part of the plasma. However, throughout this study we are always in regimes where either $T\ll m_\zprime$, or $g^\prime \ll 1$ (or both), such that the $\zprime$ induced corrections are negligible.
Likewise, gauge bosons receive an effective thermal mass which for the $\zprime$ is \cite{Bringmann:2021sth}
\begin{equation}
    m_\zprime^2 = \frac{1}{6}g^{\prime 2}T^2(N_V+N_S+\frac{N_F}{2})\,,
\end{equation}
where $N_{S,F,V}$ is the number of scalar, fermionic and vector fields in the bath and coupled to $\zprime$ and again, we are studying regions of parameter space where the thermal correction is small.

\subsection{Resonances}
Some extra care is needed for s-channel resonances, in particular if the mediator that goes on a resonance is also in thermal equilibrium, as issues of double counting arise \cite{Weldon:1983jn,Bringmann:2021sth,DeRomeri:2020wng}. To model the s-channel resonance, we adopt the Breit-Wigner form of the propagator by making the replacement
\begin{equation}
    \frac{1}{(s-m_\zprime^2)^{2}}\quad\longrightarrow \quad\frac{1}{(s-m_\zprime^2)^2+m_\zprime^2\Gamma_\zprime^2}\,.
\end{equation}

In the parameter space of interest, $\Gamma_\zprime \ll m_\zprime$ holds and it is convenient to use the narrow width approximation (NWA). In the NWA
\begin{equation}
    \frac{1}{(s-m_\zprime^2)^2+m_\zprime^2\Gamma_\zprime^2} \quad \rightarrow \quad \frac{\pi}{m_\zprime\Gamma_\zprime}\delta(s-m_\zprime^2)\,,
\end{equation}
and the on-shell production of mediators and their subsequent decay is resonantly enhanced. In case of a scalar resonance, the amplitude may be written as a sum of the resonant part with the Breit-Wigner prescription in the NWA, and the product of the amplitudes for production from inverse decays and subsequent decays. Following Reference \cite{Bringmann:2021sth}, this would correspond to 
\begin{equation}
\abs{\mathcal{M}_{f\bar f\rightarrow\nu_R\bar\nu_R}}^2 = \frac{\abs{\mathcal{M}_{f\bar f\rightarrow\zprime}}^2\abs{\mathcal{M}_{\zprime\rightarrow\nu_R\bar\nu_R}}^2}{m_\zprime\Gamma_\zprime}\pi\delta(s-m_\zprime^2)\,.
\end{equation}
For vector resonances, we have the added complication of spin correlations, and the amplitude can not be factorized exactly as above \cite{Bringmann:2021sth,Bringmann:2017sko,Uhlemann:2008pm}. However, we can introduce the decorrelated matrix element, see e.g. \cite{Uhlemann:2008pm}, to find a NWA prescription. First, we observe that for a vector in the s-channel,
\begin{equation}
\begin{aligned}
\abs{\mathcal{M}_{f\bar f\rightarrow \nu_R\bar\nu_R}}^2 &\propto  \abs{\mathcal{M}_{f\bar f\rightarrow \zprime}^\mu\left( g_{\mu\nu} - \frac{k^\mu k^\nu}{m_\zprime^2}\right)\mathcal{M}_{\zprime\rightarrow \nu_R\bar\nu_R}^\nu}^2\\ &= \abs{\sum_\lambda \mathcal{M}_{f\bar f\rightarrow \zprime}^\mu \, \varepsilon_\mu^\star(k, \lambda) \varepsilon_\nu(k, \lambda)  \, \mathcal{M}_{\zprime\rightarrow \nu_R\bar\nu_R}^\nu}^2\,.
\end{aligned}
\end{equation}
The NWA for a vector resonance corresponds to making the replacement
\begin{equation}
\begin{aligned}
    \abs{\sum_\lambda \mathcal{M}_{f\bar f\rightarrow \zprime}^\mu \, \varepsilon_\mu^\star(k, \lambda) \varepsilon_\nu(k, \lambda)  \, \mathcal{M}_{\zprime\rightarrow \nu_R\bar\nu_R}^\nu}^2
    \rightarrow  \frac{1}{3}\sum_{\lambda,\lambda^\prime} \abs{\mathcal{M}_{f\bar f\rightarrow \zprime}^\mu \, \varepsilon_\mu^\star(k, \lambda)}^2 \abs{\varepsilon_\nu(k, \lambda)  \, \mathcal{M}_{\zprime\rightarrow \nu_R\bar\nu_R}^\nu}^2\,,
    \label{eq:NWA_vector}
\end{aligned}
\end{equation}
which is correct up to neglecting spin correlations between the initial state and final state. It has been shown that for sufficiently inclusive rates such as total cross sections, spin correlations have a negligible contribution \cite{Uhlemann:2008pm}\footnote{In our case production proceeds from a thermal bath, and all possible spin configurations are assumed to be equally likely.}. We show explicitly in Appendix \ref{sec:matrix} that the exact matrix element with NWA propagator, and the decorrelated matrix element give the same resonant cross section in vacuum. Since we assume that the phase space distributions of different spin states are identical for any species, we do not expect to introduce additional correlations.

Then, as a prescription in the resonant regime, we write
\begin{equation}
\begin{aligned}
    \mathcal{C}_{f\bar f\leftrightarrow \nu_R\bar\nu_R} &= \int d\Pi_1 d\Pi_2 d\Pi_3 d\Pi_4 \, (2\pi)^4\,\delta^{(4)}(p_1+p_2-p_3-p_4)\,  \frac{\abs{\mathcal{M}_{f\bar f\rightarrow\zprime}}^2\abs{\mathcal{M}_{\zprime\rightarrow\nu_R\bar\nu_R}}^2}{2 E_\zprime m_\zprime\Gamma_\zprime}\\
    &\times\pi\delta(E_1+E_2-E_\zprime)\,\left(f_1f_2 \bar f_3 \bar f_4 - f_3 f_4 \bar f_1 \bar f_2\right)\\
    &\equiv \mathcal{C}_{f\bar f\rightarrow \nu_R\bar\nu_R} - \mathcal{C}_{f\bar f\leftarrow \nu_R\bar\nu_R} \,.
    \label{eq:resonant_coll}
\end{aligned}
\end{equation}
Here, the indices $1,2$ correspond to the initial state fermions $f,\bar f$, and indices $3,4$ denote the final state $\nu_R,\bar\nu_R$.

As was shown in reference \cite{Bringmann:2021sth}, if we we focus on the forward collision operator $\mathcal{C}_{f\bar f\rightarrow \nu_R\bar\nu_R}$ and neglect the backreaction, but assume that the mediator is in thermal equilibrium with the SM, we can write the expression as
\begin{equation}
\begin{aligned}
    \mathcal{C}_{f\bar f\rightarrow \nu_R\bar\nu_R} &= \int d\Pi_1d\Pi_2\,\frac{2\pi\,\delta^{(1)}(E_1+E_2-E_\zprime)}{4E_\zprime \Gamma_\zprime m_\zprime} \abs{\mathcal{M}_{\zprime\rightarrow f\bar f}}^2 f_1f_2\\
    &\times\int d\Pi_3 d\Pi_4 \,(2\pi)^4\, \delta^{(4)}(p_1+p_2-p_3-p_4) \abs{\mathcal{M}_{\zprime\rightarrow\nu_R\bar\nu_R}}^2 \bar f_3 \bar f_4 \\
    &=\int d\Pi_3 d\Pi_4 d\Pi_\zprime\, (2\pi)^4\,\delta^{(4)}(p_3+p_5-p_\zprime) \abs{\mathcal{M}_{\zprime\rightarrow \nu_R\bar \nu_R}}^2\, f_3f_4\bar f_\zprime \, \frac{\hat{\Gamma}_{\zprime\rightarrow f\bar f}}{\bar f_\zprime \Gamma_\zprime} \,,\\
     \label{eq:large_coll}
\end{aligned}
\end{equation}
where to arrive in the last line we expand with $\frac{\bar f_\zprime}{\bar f_\zprime}$ and complete the momentum conserving delta function for $\zprime$ with a dummy integration over $p_\zprime$. We also exploited the equilibrium assumption between $f$ and $\zprime$, which allows us to interchange $f_\zprime \bar f_1 \bar f_2 = f_1 f_2 \bar f_\zprime $. For convenience, we define the shorthand
\begin{equation}
\hat{\Gamma}_{\zprime\rightarrow f\bar f} = \frac{1}{2m_\zprime}\int d\Pi_{1}d\Pi_2\,(2\pi)^4\,\delta^{(4)}(p_1+p_2-p_\zprime)\,\abs{\mathcal{M}_{\zprime\rightarrow f\bar f}}^2 (1-f_1)\,(1-f_2)\,.
\label{eq:decay_medium}
\end{equation}
which can be interpreted as a decay width that is corrected by occupation of final states.

The general expectation is that if $f$ and $\zprime$ are in thermal equilibrium, production of $\nu_R$ should be equivalent to considering decays of $\zprime$ from the thermal bath, or in other words that Equation \ref{eq:large_coll} coincides with
\begin{equation}
    \mathcal{C}_{\zprime\rightarrow\nu_R\bar\nu_R} = \int d\Pi_1d\Pi_2 d\Pi_\zprime\, (2\pi)^4\,\delta^{(4)}(p_1+p_2-p_\zprime) \abs{\mathcal{M}_{\zprime\rightarrow f\bar f}}^2 f_1f_2 \bar f_\zprime\,
    \label{eq:decay_coll}
\end{equation}
which implies $\Gamma_\zprime = \frac{1}{\bar f_\zprime} \sum_f \hat{\Gamma}_{\zprime\rightarrow f \bar f}$ which we can understand as the appropriate Breit-Wigner width in a plasma with thermalized mediator, in accordance with the result of \cite{Bringmann:2021sth}. In the presence of non-equilibrium processes, e.g. decays into final states that are not part of the equilibrium plasma, a minor modification is necessary (see discussion in \cite{Bringmann:2021sth}). Essentially, the distribution of $\zprime$ deviates from an equilibrium distribution (its distribution is no longer set by equilibrating decays and inverse decays of plasma constituents, but also the out-of-equilibrium final states). This can be captured by adding these invisible decays to $\Gamma_\zprime$ and rescaling Equation \ref{eq:decay_coll} accordingly.

If the abundance of $\nu_R$ is set by freeze-in from the resonant contribution, but $\zprime$ is in equilibrium with the SM, it suffices to study the decay of on-shell mediators from the plasma. Since the off-resonance contribution is suppressed in higher orders of the gauge coupling, production is dominated by the resonance, or equivalently, the decay. We will study the latter for the case that $\zprime$ is thermalized, as it is particularly convenient.

If $\zprime$ is not in equilibrium with the SM, we start again from Equation \ref{eq:resonant_coll}. We perform the same computation as before, i.e. considering only the forward contribution, but this time we enforce $f_\zprime\ll 1$, as the mediator abundance is suppressed, and readily arrive at
\begin{equation}
\begin{aligned}
    \mathcal{C}_{f\bar f\rightarrow \nu_R\bar\nu_R} &= \int d\Pi_1 d\Pi_2 d\Pi_\zprime \,(2\pi)^4\,\delta^{(4)}(p_1+p_2-p_\zprime)\,\abs{\mathcal{M}_{f\bar f\rightarrow \zprime}}^2\, \hat{\text{Br}}_{\zprime\rightarrow\nu_R\bar\nu_R}\,.
    \label{eq:coll_non_thermal}
\end{aligned}
\end{equation}
Again, the hat indicates that  $\hat{\mathrm{Br}}_{\zprime\rightarrow\nu_R\bar\nu_R}$ is the branching ration in medium that includes final state occupations.

Neglecting the final state statistics remains correct as far as the frozen-in species is concerned explicitly. However, in the case of resonant production without a mediator in equilibrium, the production has a dependence on the branching ratio in medium -- which is affected by the final state occupation of SM particles and constitutes a correction due to final state statistics. Thus, we observe an indirect final state dependence for the frozen-in species from the medium dependent decay width into SM fermions which can have profound consequences, however in the situations we consider this effect is expected to be negligibly small (see also discussion in Reference \cite{Belanger:2018ccd}). 

To summarize, in the regime where the mediator is thermalized and the explicit $2\rightarrow 2$ contribution is suppressed compared to the resonant part, we study decays instead of the $2\rightarrow 2$ process, and we avoid the issue of double counting. If the mediator is no longer thermalized, we use Equation \ref{eq:coll_non_thermal} for freeze-in production, as the abundance of $\nu_R$ is also suppressed. We see here that such a prescription explicitly arises from treating the resonant part of the  $2\rightarrow 2$, while bearing in mind that the production occurs in a plasma background, thus making the possible double counting issue manifest.

Lastly, we highlight that the approach and the numerical scheme we develop here and in Appendix \ref{sec:integrations} are general, and can easily be adapted to other systems of integrated Boltzmann equations, e.g. dark matter relic abundance calculations, and provide an alternative to either other approximate techniques or rather involved complete inclusions of final state effects. The ease of implementation compared to exact reduced integral expression as in e.g. \cite{Arcadi:2019oxh,Belanger:2018ccd,Bringmann:2021sth,Lebedev:2019ton} would make such an approach lucrative if the problem is not limited by computational ressources. Moreover, extending the numerical integration beyond $2\rightarrow 2$ processes is straightforward and we can keep the consistent inclusion of relativistic and spin-statistic effects in this case.

\section{Results}
\label{sec:results}
\subsection{Standard cosmology}
Before discussing the results in more detail, let us briefly stress again the explicit and implicit assumptions we made. Neutrinos are Dirac in nature, and their Dirac nature is in particular protected by $U(1)_{B-L}$, implying that if this symmetry is broken, $\Delta (B-L) \neq 2$. Moreover, we have an implicit assumption on the reheating temperature. We either consider thermalized $\nu_R$ that leaves equilibrium at some point, or freeze-in production which peaks at $T\sim m_\zprime/3$. In either case, for the result to be applicable we need the Universe to reheat to $T_\text{reh}\sim m_\zprime$, as the bulk of $\nu_R$ production will occur for temperatures smaller than the mass.

We show the improved limits and forecasts in Figure \ref{fig:limits}. The current experimental bound of $\dneff \simeq 0.17$ implies that if the three $\nu_R$ have been in thermal equilibrium with the SM, they must have decoupled at a temperature of about $T_\text{dec}\sim 40\mathrm{\,GeV}$ (see Equation \ref{eq:basis}). For the freeze-out limit, we compared the result of our exact computation including thermal masses with simplified prescriptions for the backreaction term \cite{Adshead:2020ekg, Luo:2020fdt}, no final state statistics, and vacuum masses. The corrections we find are on a percent level.

If thermal equilibrium has never been established, the relic abundance will be set by freeze-in production which peaks around $T\sim m_\zprime/3$. The qualitative behavior of the updated limits (black curve in Figure \ref{fig:limits}) is analogous to previous studies \cite{Abazajian:2019oqj,Adshead:2022ovo} for larger values of $\dneff$. For mediator masses $m_\zprime \gg 1\mathrm{\,TeV}$, the decoupling temperature is below the resonant regime $T\lesssim m_\zprime$, and so for the abundance to saturate the current limits on $\dneff$, the $\nu_R$ inevitably thermalize at high temperatures and the relic abundance is subsequently determined by the decoupling process near $T\sim 40\mathrm{\,GeV}$.
As we transition to smaller masses, the peak of resonant production is similar to or falls below the estimated decoupling temperature. Consequently, the relic abundance cannot be set by decoupling alone anymore, but is given predominantly by the freeze-in from inverse decays at lower temperatures. Due to resonant enhancement, the limits become significantly stronger, as we fully transition to the freeze-in regime. 

We also provide a refinement on projected experimental reaches. Here we focus on the expected sensitivity of the original proposal for CMB-S4 ($\dneff < 0.06$), and also compare to previously determined forecasts from \cite{Adshead:2022ovo}, which approximates final state statistics and the backreaction term. In addition, we include thermal masses as a leading thermal correction.
As discussed in Section \ref{sec:formalism}, the out of equilibrium production of mediators becomes indirectly sensitive to final state statistics.  While decays into $\nu_R$ are not Pauli-blocked, decay to SM fermions are to some extent. Consequently, the branching ratio receives an in medium enhancement, production of $\nu_R$ is boosted. The effect is small and combined with thermal masses the corrections we find generally do not effect results beyond percent level, which is consistent with uncertainty estimates.

We note also that, as we approach smaller values of the mass and small gauge couplings, the total decay width becomes comparable to or smaller than the Hubble rate, $\Gamma_\zprime\lesssim H$, and the $\zprime$ becomes cosmologically long-lived. After production, they redshift as matter and slower than the ambient radiation bath. Thus, the energy injection at decay is larger compared to prompt decays close to production \cite{Li:2023puz,Adshead:2022ovo,Esseili:2023ldf}. We have also considered this explicit production of $\zprime$ by explicitly tracking their evolution as non-relativistic matter before decaying into SM final states and right-handed neutrinos. For the masses and couplings we consider, the effect is negligibly small and typically on a sub percent level, in accordance with the estimates from \cite{Adshead:2022ovo}. Nevertheless, we note a disagreement of up to $\sim10\%$ above $m_\zprime\sim 1\mathrm{\,GeV}$ that also persists if we make the same approximations as \cite{Adshead:2022ovo} and consistently include the delayed decay. We therefore attribute this difference to numerical difficulties.

Forecasted limits we provide can be extrapolated to in principle arbitrarily high gauge boson masses. Indeed, for $m_\zprime\sim \Lambda_\text{GUT}\sim 10^{16}\mathrm{\,GeV}$ with $g^\prime\sim \mathcal{O}(1)$ is at the edge of what seems to be detectable in the near future. In turn this would imply that any natural   $g^\prime\sim \mathcal{O}(1)$ realization below the GUT scale is accessible to us, and absence of detection would pose serious problems for large classes of models that preserves the Dirac nature of the neutrino by means of a gauge symmetry. We note here that such a conclusion makes strong assumptions on the thermal history. We revisit and relax these assumptions in the next section.

The updated limits and forecasts presented here will also hold as good approximations to other $U(1)$ extensions of the SM. Prerequisite is the existence of light $\nu_R$, and that $\nu_R$ and at least a subset of SM fermions are subject to the the same new gauge interaction. Then, approximate limits on such models may be obtained by making an appropriate rescaling of gauge couplings.

\subsection{Non-standard cosmologies}
For now, let us assume that the Universe has reheated to the aforementioned $T\sim m_\zprime$ or higher, and that the abundance of right-handed neutrinos has been set. If couplings are large, and $\nu_R$ remains coupled to the SM, then whatever change we make to the SM plasma will propagate also to the $\nu_R$ fluid. Only changes to the SM plasma after decoupling of the right-handed neutrinos will affect the ratio $\rho_{\nu_R}/\rho_\SM$, and thus $\dneff$.
Of course it is feasible that $\nu_R$-genesis falls into a non-standard episode of the cosmological history. While we do not study this case in detail, we argue that for many scenarios such a modified history during production is typically accompanied by a period of reheating of the SM. Thus, the relic abundance from this period is diluted, and the dominant part is given by subsequent production after reheating.

As we argue, the most significant changes to the naive predictions of $\dneff$ come from entropy injection after the initial $\nu_R$ abundance is set. Then, $\dneff\propto \rho_{\nu_R}/\rho_\SM$, and $\rho \propto s^{4/3}$\footnote{If we are for the moment agnostic about the subtle differences in the relativistic degrees of freedom.}. We define a dilution factor $D=S_\text{after}/S_\text{before}$, and this definition implies
\begin{equation}
    \dneff \,\rightarrow \,\dneff \,\times\, \left(\frac{D_{s,\nu_R}}{D_{s,\SM}}\right)^{4/3}\,.
    \label{eq:dneff_rescaling}
\end{equation}
In most cases, there is negligible entropy injection into the $\nu_R$ sector, and so $D_{s,\nu_R}=1$. Equation \ref{eq:dneff_rescaling} provides a convenient way to rescale limits we find when there is additional entropy being produced. We provide approximate expressions for the dilution factors in various modified thermal histories, which enables the recasting of our limits.

We can summarize a large class of modified cosmologies by introducing an additional fiducial energy density $\rho_\phi$. To this end we augment Equation \ref{eq:energy_density}
\begin{align}
    \frac{d\rho_\phi}{dt} + 3H(1+w_\phi) \rho_\phi &= -\Gamma_\phi \rho_\phi \\
    \frac{d\rho_\SM}{dt} + 3H(\rho_\SM + P_\SM) &= (1-\kappa)\Gamma_\phi \rho_\phi - \mathcal{C}_{f\bar f\rightarrow \nu_R \bar \nu_R} \\
    \frac{d\rho_{\nu_R}}{dt} + 4H\rho_{\nu_R} &=  \,\mathcal{C}_{f\bar f\rightarrow \nu_R \bar \nu_R} + \kappa \,\Gamma_\phi \rho_\phi\,.
\end{align}
The equation of state $w_\phi$, the decay rate $\Gamma_\phi$ and the relative fractions injected into the respective species fully describe the modified cosmology. In particular, it contains variations of dark energies ($w_\phi= -1$), non-relativistic species or coherent oscillations with a potential phase of early matter domination ($w_\phi=0$), or additional dark radiation ($w_\phi = 1/3$). Starting from this parametrization, we provide dilution factors for selected benchmark scenarios.

\subsubsection{Extra degrees of freedom}
The first situation we consider is the presence of additional thermalized degrees of freedom, collectively labeled as $\phi$ here. In the language of our augmented cosmological fluid equations, we consider any number of extra degrees of freedom captured in $\rho_\phi$ as radiation-like and tightly coupled to the SM plasma. Then, we need to study $\nu_R$ production relative to the $\SM+\phi$ plasma. Every degree of freedom freezing out from the plasma increases the SM temperature courtesy of entropy conservation. The dilution factor amounts to
\begin{equation}
    D_s = \left(\frac{g^{(s)}_\SM+g^{(s)}_\phi}{g^{(s)}_\SM} \right)\,.
\end{equation}
Even if we double the degrees of freedom, which is roughly what we would expect e.g. from Supersymmetry, we only reduce $\dneff$ by a little more than half. We conclude that additional degrees of freedom between the electroweak scale and end of $\nu_R$ production, unless very numerous, do not significantly reduce the prospects of detection.

\subsubsection{Phase transitions and other vacuum energies}
We consider a vacuum like component of energy that eventually decays via an intermediate stage that redshifts as matter. This parametrization captures many models of early dark energies, but also phase transitions in the early Universe. We largely follow the analytic estimates from \cite{Benso:2025vgm}, and only make the substitutions necessary for our discussion.
We define $T_i$($a_i$) as the temperature (scale factor) at which $\phi$ starts to dominate, i.e. $\rho_\phi=\rho_\SM$. The entropy injection phase is characterized by the non-adiabatic evolution of the SM plasma and starts at $T_e$ when noticeable amounts of energy are transferred to the SM bath. Eventually, we once more reach an adiabatic radiation-like SM at $T_r$ which we define approximately from the time at which $\Gamma_\phi = H$ holds. Instead, if the decay is triggered at some $T_e$ and $\Gamma_\phi \gg H(T_e)$ holds, the vacuum energy decays very efficiently. Energy is promptly transferred to the respective sectors. Taking results from \cite{Benso:2025vgm}, we omit a possible entropy transfer to the $\nu_R$ sector ($\kappa = 0$) and we find for the instantaneous entropy injection
\begin{equation}
    D_s = \frac{T_r^3}{T_e^3} = \left(1 + \frac{30\Delta V}{\pi^2 g_{(\rho)} T_e^4}\right)^{3/4} 
    \simeq \left(1+\alpha\right)^{3/4}\,,
\end{equation}
where $\alpha = \Delta V/\rho_\SM(T_e)$. Thus, only significant supercooling in a phase transition, or a period of significant vacuum energy domination with subsequent reheating could cause a meaningful dilution of $\nu_R$.

The case of an extended decay phase is slightly more involved. Again, adapting the solution from \cite{Benso:2025vgm}, the dilution factor if the vacuum energy dominates prior to its decay, in this case reads
\begin{equation}
    D_s = \left(\frac{90}{8\pi^3g_{(\rho)}(T_r)}\right)^{3/4} \left(\frac{2}{5}\Gamma_\phi M_P\right)^{1/2} \left(\frac{8\pi \Delta V}{3}\right) \frac{1}{T_e^3}\,,
\end{equation}
which e.g. applies to a phase transition, where the energy density transfers into oscillations of the scalar field around the new minimum and the subsequent decay of the oscillations is slow. In most cases, the prompt injection estimate gives a reasonable indication if reheating of the SM is significant.

Thus, in order for a phase transition or other forms of early dark energy to have a meaningful impact on primordial $\nu_R$-genesis, they need to be the dominant energy density prior to reheating. Only then is a dilution of $\nu_R$ relative to the SM that exceeds $\mathcal{O}(1)$ feasible from sufficient reheating. Should this occur, we can just consider the production of $\nu_R$ after reheating instead and the limits that cast on $\nu_R$ are consequently weakened.

Notably, a BSM induced first order electroweak phase transition or modified QCD transitions are often expected to fall within the $\alpha < 1$ regime, so modifications to these transitions will not challenge the robustness of our limits (see e.g. the discussion of models in \cite{Caprini:2019egz}).

\subsubsection{Early matter dominated era}
We sketch the changes introduced to cosmology by an era of early matter domination (EMD), by adopting an appropriate phenomenological prescription. Thus, we are agnostic to whether EMD is induced by the presence of meta-stable heavy particle \cite{Allahverdi:2021grt}, string moduli or other sources of EMD.
The matter era can fully be characterized by defining $T_\text{eq}$ as the temperature, at which $\rho_\phi=\rho_\SM$. The temperature at which significant energy is injected into the plasma is denoted by $T_e$. This injection is characterized by a phase of non-adiabatic evolution for the SM plasma, and lastly $T_r$ is defined by $\Gamma_\phi = H(T_r)$, which is approximately the time at which the reheated plasma evolves adiabatically once more.

During the matter dominated era the Hubble rate decays slower and it is straightforward to find $H\propto a^{-3} \propto T^{3/2}$. Both the possible $\nu_R$ population as well as the SM plasma continue to evolve adiabatically, i.e. $\rho \propto a^{-4}$. Consequently, their ratio remains unaffected if $\nu_R$-genesis took place before the onset of the matter dominated era. Production during such an era will be different. For example, for a freeze-in like production, the ratio $\mathcal{C}/H$ is important, and due to the slowed down decay of the Hubble rate may be affected. We do not study this case in more detail, since any meaningful period of early matter domination with subsequent reheating will usually dilute such a primordial $\nu_R$ abundance. We note that this type of production has been studied extensively for dark matter relic abundances (e.g. \cite{Drees:2017iod}), and at least for UV-sensitive freeze-in, production from this regime is found to be negligible \cite{Benso:2025vgm}.

During the entropy injection phase $T\propto g_{(\rho)}^{-1/4} a^{-3/8}$ holds, which indicates that entropy is injected into the SM plasma. Assuming negligible changes for $\nu_R$, a dilution factor can be inferred \cite{Dutra:2021phm}
\begin{equation}
    D_s = \frac{T_\text{eq}}{T_r} \frac{g_{(\rho)}(T_\text{eq})  g_{(s)}(T_r) }{g_{(\rho)}(T_r) g_{(s)}(T_\text{eq})} \simeq \frac{T_\text{eq}}{T_r}\,.
\end{equation}
Thus, dilution is approximately given by the ratio of temperatures between the onset of the matter dominated era, and its end, and unless the matter dominated era was very brief\footnote{Here we gloss over the fact that for a very brief matter era, the approximations needed for this result will break down. Our conclusions are not affected by this.}, dilution generically yields $D_s \gg 1$. Therefore, we become practically insensitive to primordial abundances for typical periods of extended matter domination. Only production in the last stages of the entropy injection phase may leave an imprint, similar to our discussion for the vacuum dominated Universe. As a conservative estimate, production starting from the second reheating temperature $T_r$ can be used.

\subsubsection{Low reheating temperature}

\begin{figure}
    \centering
    \includegraphics[width=0.49\linewidth]{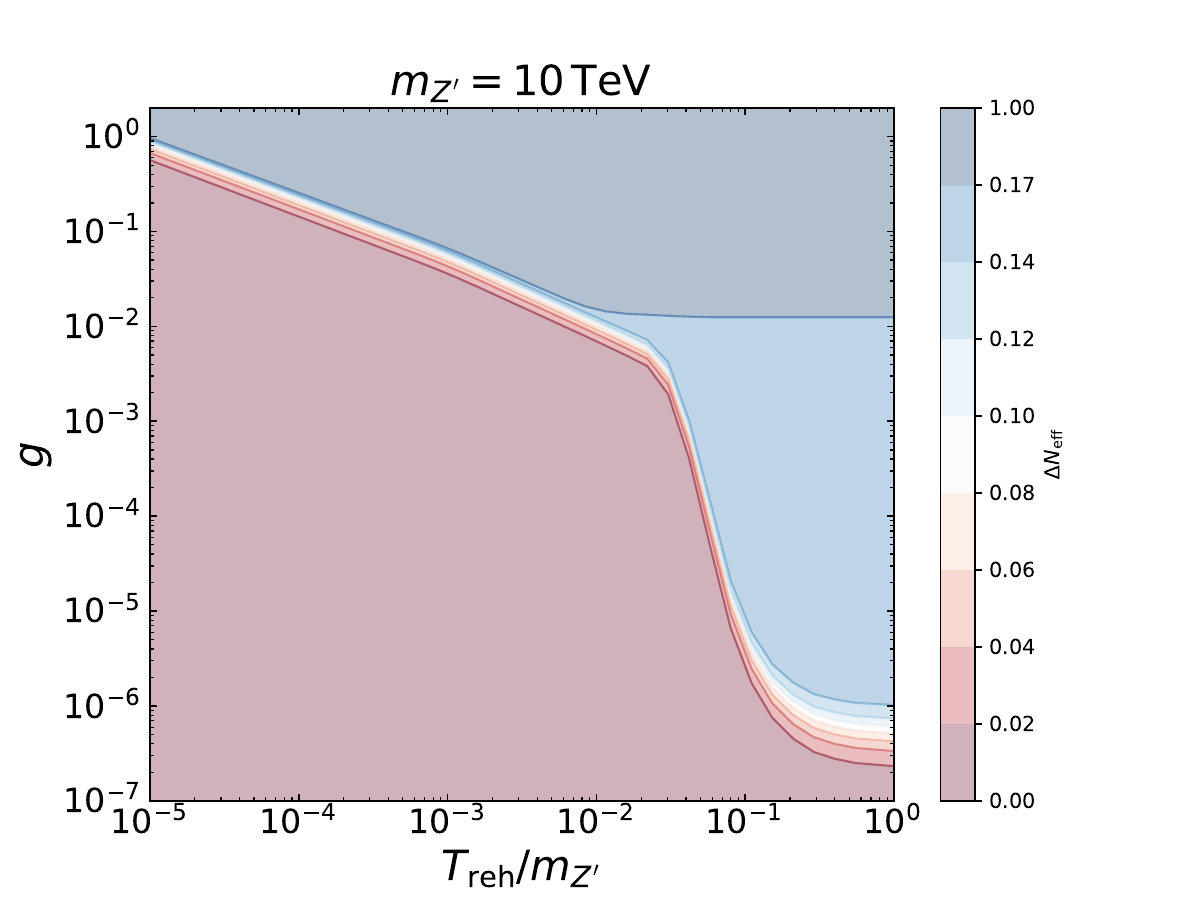}
    \includegraphics[width=0.49\linewidth]{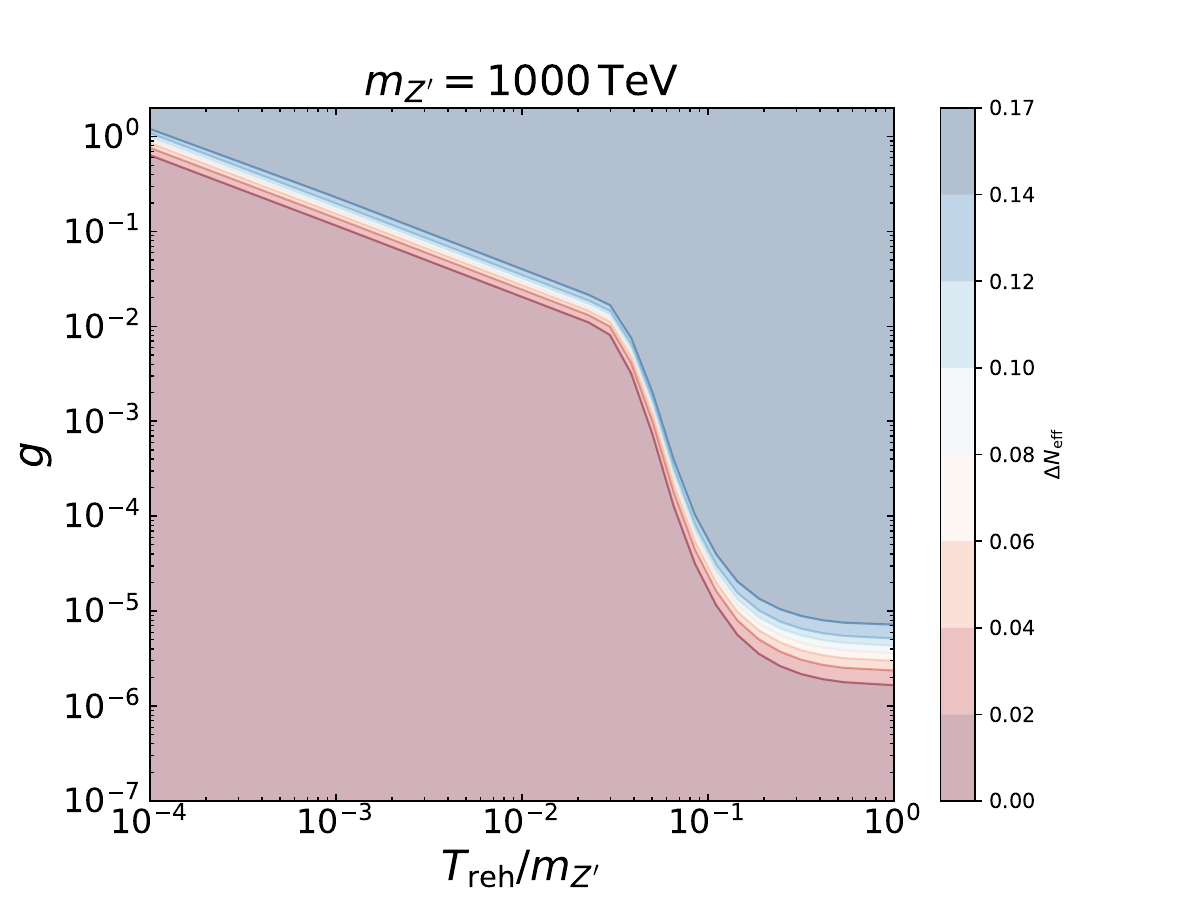}
    \includegraphics[width=0.49\linewidth]{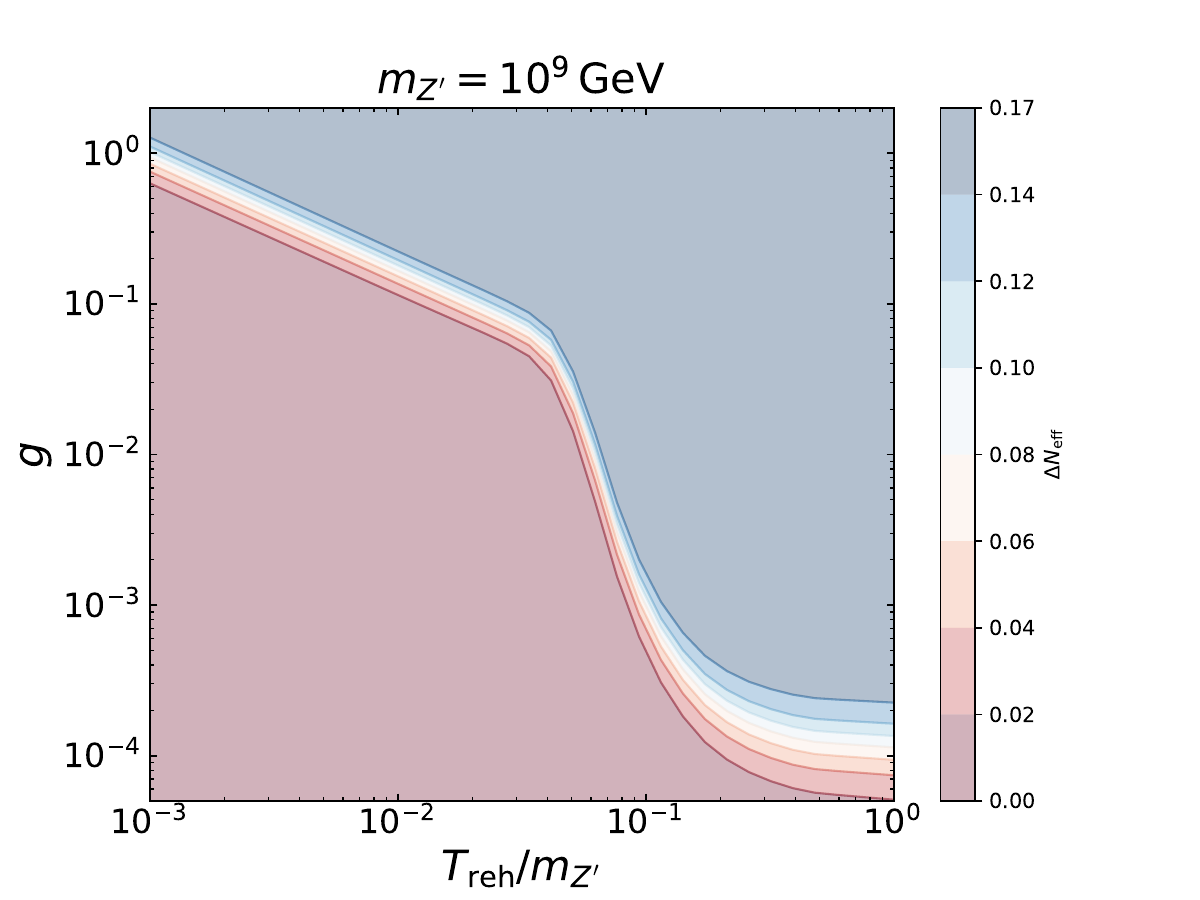}
    \includegraphics[width=0.49\linewidth]{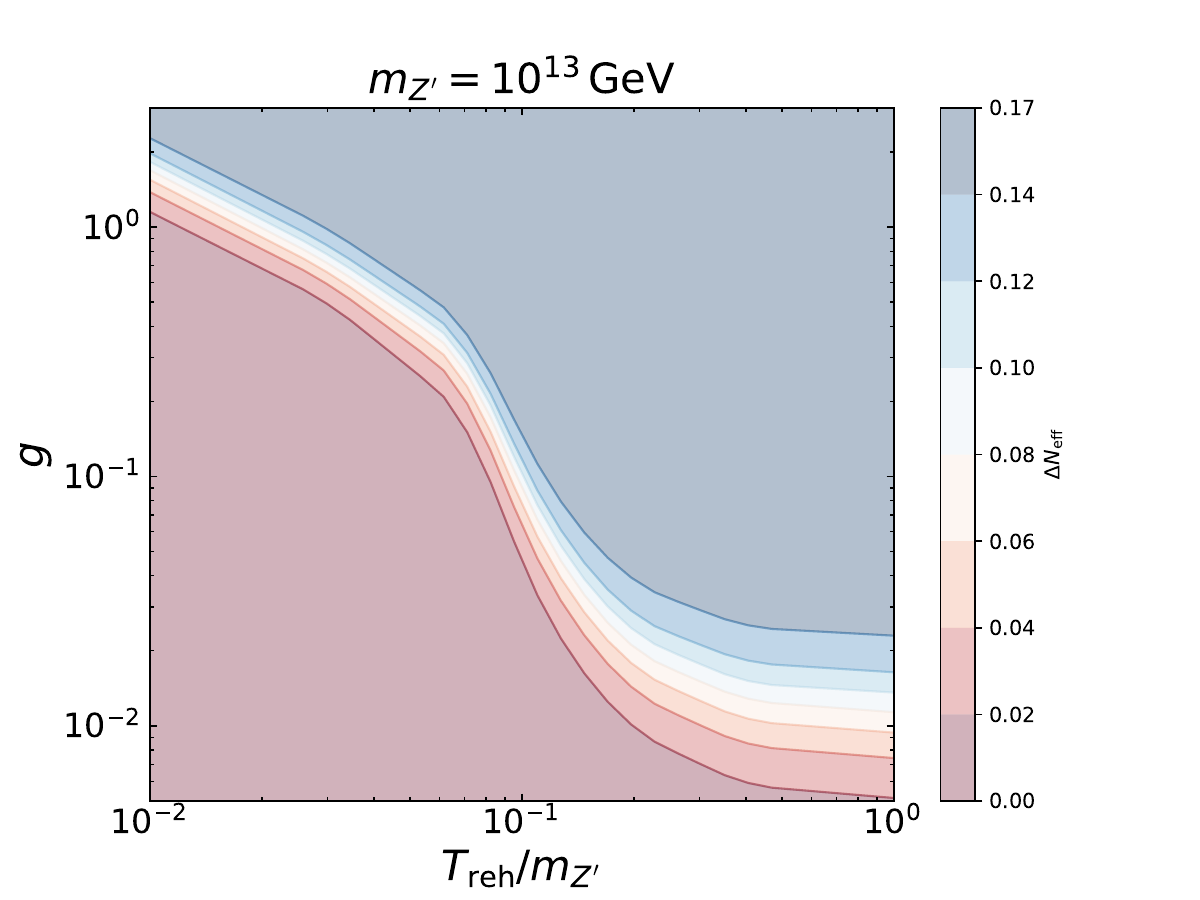}
    \includegraphics[width=0.49\linewidth]{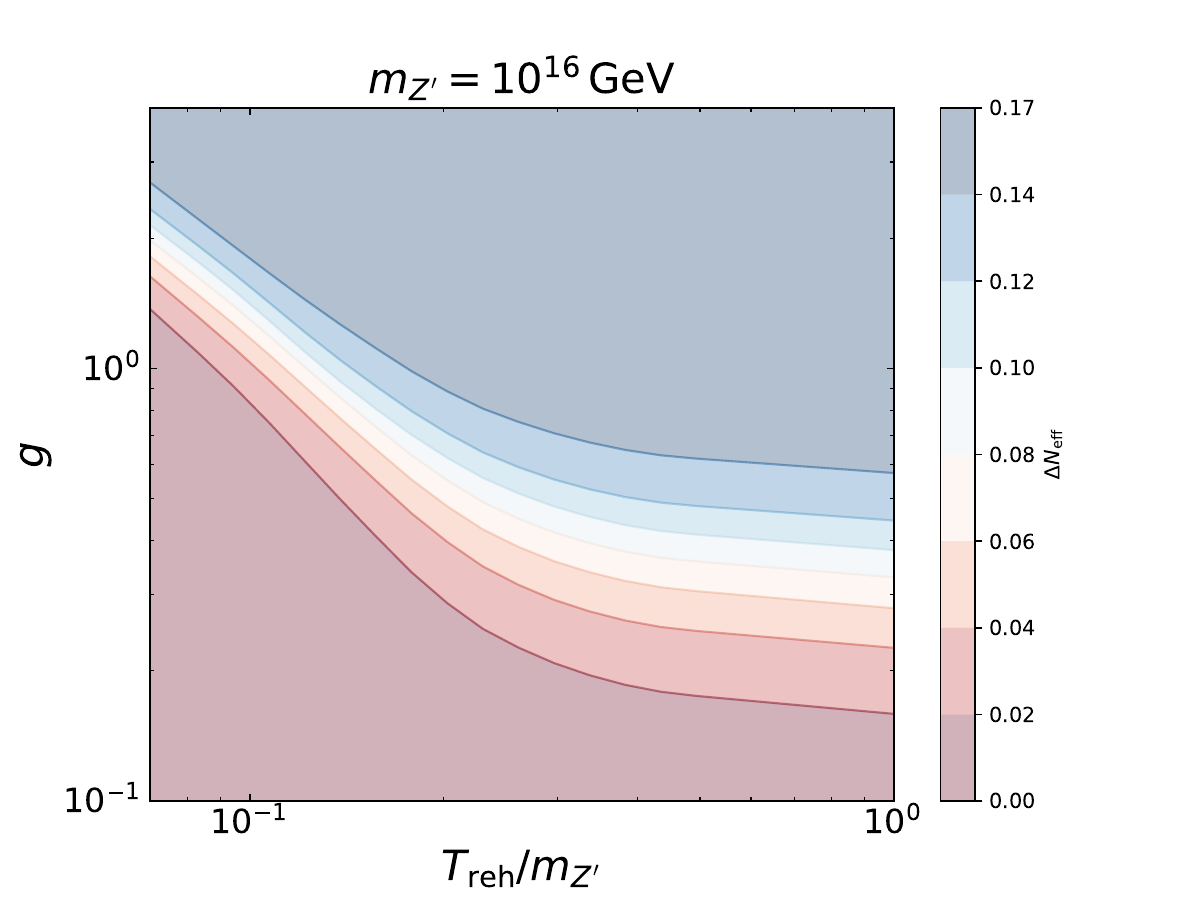}
    \caption{The effect of a lowered reheating temperature for selected values of the $\zprime$ mass. For masses up to thousands of $\mathrm{TeV}$, limits can be avoided by having a significantly lowered reheating temperature only. Especially for $m_\zprime\sim\mathrm{TeV}$ and $g \sim \mathcal{O}(1)$ we can still have a $\zprime$ within collider reach, if there was a significant reheating event as low as temperatures $T_\text{reh} \lesssim 100\mathrm{\,MeV}$. For larger masses, strong limits on $\mathcal{O}(1)$ gauge coupling may be avoided by having (inflationary or secondary) reheating at temperatures a few orders of magnitude below the gauge boson mass. We note that for scales as high as the GUT scale, if the Universe actually reheated to such high temperatures, a seemingly natural realization with order 1 gauge coupling becomes testable. However, reheating to just an order of magnitude below the mass will suppress any signal from $\dneff$, and GUT scale $\zprime$ with Dirac neutrinos may still avoid. We note that the necessary reheating temperature to produce detectable amounts of $\nu_R$ at the GUT scale is in tension with upper limits on the scale of inflation.}
    \label{fig:reheating}
\end{figure}

The last case we consider is that of a significantly lowered reheating temperature. This is relevant for two reasons: First, an outright low reheating temperature coming from inflation is possible. Afterall, the most stringent bounds on the reheating temperature come from not spoiling the success of BBN at $T_\text{BBN} \sim \mathrm{MeV}$. Moreover, from the previous considerations we infer that many modified cosmologies lead to a (partial) reheating at lower temperatures. As far as primordial $\nu_R$ abundances are concerned, even a rather mild reheating can significantly suppress $\dneff$ detection prospects, so studying production from $T_\text{reh}<m_\zprime$ is important.

For masses relevant in collider searches, this has also been considered in \cite{Caloni:2024olo}, but we extend this to the entire mass range up to the GUT scale and we utilize our improved calculation scheme for $\nu_R$-genesis. To mimic the effect of low reheating, we start integrating the relevant equations not from $T\gg m_\zprime$ as in the previous section, but rather from a chosen value $T_\text{reh}<m_\zprime$. Doing this simplifies the problem considerably, as we omit the challenging and chaotic phase of reheating and possible $\nu_R$ produced in the reheating process itself. In that sense, the results we provide are conservative, as we capture only production after reheating completed. In particular, direct reheating contributions, and possible primordial relics that have not fully diluted away in the reheating process are neglected.

We show the results from reduced reheating temperatures in Figure \ref{fig:reheating} for selected values of $m_\zprime$. We focus on near- and far-future collider reaches, and selected values between the anticipated futuristic collider reaches and the GUT scale. Limits remain stringent for reheating close to the mass of the gauge boson, as this regime is dominated by the resonantly enhanced inverse decay and subsequent decays of $\zprime$. We then observe a weakening in limits, as the on-shell production becomes kinematically impossible after reheating. For even lower reheating temperatures of $T_\text{reh} \lesssim 10^{-2}\,m_\zprime$, the relation between gauge coupling and reheating temperature becomes universal.
This is the UV-sensitive freeze-in of $f\bar f\rightarrow \nu_R \bar \nu_R$ via an effective Four-Fermi interaction. Production is determined by the highest temperature reached, and relic energy densities in this regime are controlled approximately by the combination $ g^4 T_\text{reh}^3 M_P/m_\zprime^4$. We find for a benchmark value $\dneff \sim 0.1$ the approximate empirical relations
\begin{align}
    g &\sim 1.8 \,\, \left(\frac{T_\text{reh}/m_\zprime}{10^{-3}}\right)^{-3/4} \left(\frac{m_\zprime}{10^9 \mathrm{\, GeV}}\right)\,, \\
    \frac{T_\text{reh}}{m_\zprime} &\sim 0.05 \left(\frac{m_\zprime}{10^9\mathrm{\,GeV}}\right)^{1/3} \left(\frac{g^\prime}{0.1}\right)^{-4/3}\,.
\end{align}
These convenience expressions give a good first estimate as to an extended model can evade bounds if the reheating temperature is lowered.

We observe that, in particular for extremely heavy $\zprime$, the strong limits we would get from extrapolating the forecasts in Figure \ref{fig:limits} can be weakened substantially. For scales as high as the GUT scale, we observe that a testable $\dneff$ contribution requires reheating temperatures close to the upper bounds on the scale of inflation \cite{Planck:2018jri}. Below that, strong bounds on on heavy $\zprime$ with $\mathcal{O}(1)$ gauge couplings can be relaxed if reheating is only two to three orders of magnitude below the gauge boson mass. As we progress to smaller masses in the $10-1000\mathrm{\,TeV}$ range, evading current and projected limits for natural choices of gauge coupling requires going to very small reheating temperatures significantly below the temperature of EWSB. These ranges could in principle be considered testable in the laboratory, and so the possible implications on the reheating temperature are particularly interesting to consider here. 

This is especially interesting if we consider our minimal model as a building block of more complex models that aim to incorporate other phenomena, such as including suitable dark matter candidates. The irreducible $\nu_R$ contribution force such models to potentially low reheating temperatures, with implications for dark matter production, phase transitions, or other cosmological observables.
We note here in particular possible consequences for Dirac leptogenesis \cite{Dick:1999je}. The requirement that $\nu_R$ and the SM are decoupled during the era in which sphalerons are active implies that future experiments will strongly constrain the feasibility of Dirac leptogenesis in models that include gauged $B-L$ or similar $\zprime$ extensions. Unless it seems desirable to go to sufficiently small gauge coupling, such bounds on leptogenesis can again only be avoided by lowering the reheating temperature -- which in return narrows the window on creating the initial lepton asymmetry and the efficiency of the sphaleron process in models with Dirac leptogenesis considerably.

\section{Conclusion}
\label{sec:conclusions}
Models that predict neutrinos to be Dirac particles remain well motivated and widely studied. Such models often include new gauged $U(1)$ symmetries, and the combination of a $\zprime$ with Dirac neutrinos is strongly constrained by cosmology. In particular, gauged extension provide an explanation of the Dirac nature of the neutrino, and among possible extensions, a gauged $U(1)_{B-L}$ provides a useful benchmark case. The gauge symmetry protects Dirac nature of the light neutrinos even if broken, as long as $\Delta(B-L)\neq 2$.

The model we adopt in this study is minimal, but arises as a building block in many non-minimal realizations of gauged $U(1)_{B-L}$ with Dirac neutrinos, or similar structures for more general $\zprime$ extensions. We then compute contributions of the light-right handed neutrino to the total radiation budget of the Universe.
We sketch and employ an integration scheme for evaluating collision operators, as they appear in integrated Boltzmann equations without any additional assumptions. The scheme proves to be computational reliable and efficient, and in principle enables convenient implementations of a variety of processes. Notably, we can compute results without making simplifying assumptions on final state statistics or the backreaction, but only have to adopt a prescription for the underlying phase space density, which in many situations can be physically motivated. While for this study these effects do not give rise to significant corrections, when applied to different problems our scheme will make the inclusion of such effects relatively easy.

Using the new constraint on the extra radiation $\dneff<0.17$ ($95\%$ C.L.) from the latest ACT DR6 data release, we update the respective limits in the $m_\zprime-g^\prime$ plane. The new limits are orders of magnitude stronger than previous cosmological limits coming from BBN and Planck, and surpass collider limits by a similar margin and on all scales we can test for now. Even the forecasted reach of the FCC is barely competitive compared to the updated cosmological limit, and also only for a narrow range of gauge boson masses.  Notably, we provide for the first time limits in the otherwise inaccessible range of $m_\zprime > 4\mathrm{\,TeV}$, and excluding gauge boson masses of $m_\zprime \sim 100\mathrm{\,TeV}$ for $g^\prime=\mathcal{O}(1)$.

We show the potential of future CMB experiments by quoting results for a benchmark of $\dneff=0.06$ and find our forecasts to be consistent with previous results. We provide exact solutions to the integrated Boltzmann equations and find that corrections to simplified approaches are generically of $\mathcal{O}(1\%)$, which is consistent with previous uncertainty estimates. When extrapolated to higher masses, the forecasts suggest that even GUT scale realizations of $B-L$ with $g^\prime\sim \mathcal{O}(1)$ are testable, although careful examination of the underlying assumptions makes this unlikely. However, natural realizations of gauged $U(1)_{B-L}$ with Dirac neutrinos will be severely constrained by future observations on all scales up to $\Lambda_\text{GUT}$ within a standard thermal history.

The limits we find depend on assumptions on the underlying cosmology, namely that the Universe reached temperatures $T\sim m_\zprime$ and higher, and that the thermal history proceeded to be radiation-like and without major entropy generating events between creation of $\nu_R$ and BBN. We discuss at length the effect of modifications to this assumptions. Notably, we consider the presence of extra degrees of freedom in the plasma that eventually freeze-out, as well as entropy injection coming from phase transitions, early dark energies, and a period of early matter domination. We provide recipes to translate our bounds to scenarios of a modified thermal history. Constraints are robust when facing typical non-standard electroweak or QCD phase transitions, but note that a period of early matter domination generically leads to significant reheating that will dilute a relic $\nu_R$ beyond detection. The dominant contribution is expected to come from the UV-sensitive, post-reheating production of $\nu_R$ in those scenarios.

We consider the effect of lowered reheating temperatures in detail, providing modified limits for selected benchmark points and show that constraints can be partially evaded if the Universe initially reheats to $T_\text{reh} \ll m_\zprime$ or faces a second period of reheating, e.g. from an early matter era. In particular, scenarios close to the GUT scale that are seemingly testable in the near future when extrapolating our limits are shown to evade constraints, mostly because the reheating temperatures required are in tension with upper bounds on the reheating temperature from inflation. Taking reheating into account, natural realizations $g\sim \mathcal{O}(1)$ across all mass scales will evade cosmological limits only when the reheating temperature is lowered substantially. This provides additional constraints on extended models that require reaching certain temperatures to achieve dark matter production or Dirac leptogenesis. In particular, finding a $\zprime$ coupled to Dirac neutrinos at current or future colliders would imply a profound deviation from the early radiation dominated Universe.

An important result is that for the first time, cosmological bounds exceed laboratory and collider bounds by a significant margin on all mass scales. Previously unconstrained mass scales in the hundreds of $\mathrm{TeV}$ regime are now being tested and even higer scales seem within reach. Future experiments have the ability to provide tremendous improvements on limits on Dirac neutrinos subject to a shared gauge interaction with SM particles. The limits are generic and apply to a wide class of models that involve a gauged $B-L$ (and with small modifications other generic gauged $U(1)$ extensions) as a building block, assuming neutrinos are Dirac particles. Should future experiments prevail and provide their forecasted limits without any clear sign of excess radiation, it would provide a strong indication towards the viability of Dirac neutrino models. Likewise, if the Dirac nature of neutrinos were be established by other means, we face a strong hint towards a non-standard thermal history of the early Universe.

\begin{acknowledgments}
The authors thank Salvador Centelles Chuli\'a, Thede de Boer, and Yi Chung for helpful discussions and Paul Frederik Depta for valuable comments on the numerical implementation. TH acknowledges support by the IMPRS-PTFS. This research work made extensive use of \texttt{NumPy}\cite{harris2020array}, \texttt{SciPy}\cite{2020SciPy-NMeth} and \texttt{Matplotlib}\cite{Hunter:2007}
\end{acknowledgments}

\appendix

\section{Integrating collision operators}
\label{sec:integrations}
We compute the integrals of the collision operator using a Monte Carlo approach. To this end, we utilize the \texttt{VEGAS} framework \cite{PETERLEPAGE1978192,Lepage:2020tgj}, which through adaptive sampling achieves fast convergence of integrals and enables us to integrate collision operators efficiently. In the following, we show the implementation steps of $2\rightarrow 2$ and the $1\rightarrow 2$ collision operators. We comment briefly on subsequent implementation steps, and hope that the steps and comments presented here prove useful to future implementations of similar strategies.

\subsection{$2\rightarrow 2$ processes}
We reduce this collision operator from $12$ to $5$ dimensions, similar to \cite{Bringmann:2022aim} although we deviate for some steps.
We start with an integral of an arbitrary function of the momenta over the 4-particle phase space
\begin{align}
    \mathcal{C}_{(F)} =& \int \dps{1}\dps{2}\dps{3}\dps{4} (2\pi)^4 \delta^{(4)}\left(p_1+p_2-p_3-p_4\right)\, F(p_i)\,.
\end{align}

It is always possible go to a coordinate system in which $\vec p_1=(0,0,p_1)^T$. Angular integrations of the first momentum are then trivial and we replace $d^3p_1 = 4\pi p_1^2 dp_1$. Then, $\vec p_2 = p_2 (\sin\beta,0,\cos\beta)^T$, and $\vec p_3 = p_3 (\sin\theta\cos\phi,\sin\theta\sin\phi,\cos\theta)^T$, with respect to the fixed direction of $\vec p_1$.
Performing the trivial  azimuthal integration of $p_2$ yields
\begin{align}
    C_{(F)} =&  2(2\pi)^2 \int_0^\infty \frac{dp_1 p_1^2}{(2\pi)^3 2 E_1}
    \int_{-1}^1 d\cos\beta \int_0^\infty \frac{dp_2 p_2^2}{(2\pi)^3 2 E_2}
    \int_0^{2\pi} d\phi \int_{-1}^{1}d\cos\theta \\
    &\int_0^\infty \frac{dp_3 p_3^2}{(2\pi)^3 2 E_3} \int \dps{4} 
    (2\pi)^4 \delta^{(4)}\left(p_1+p_2-p_3-p_4\right) \times F(p_i)\,.
\end{align}
We continue by eliminating the spatial part of the $\delta$-distribution by integrating $p_4$, leaving $\delta(E_1+E_2-E_3-E_4)$. Since spatial momentum conservation has been enforced, we understand $E_4$ as a shorthand for
\begin{align}
E_4 =& \sqrt{m_4^2+(\vec p_1+ \vec p_2 -\vec p_4)^2} \\=& (m_4^2 + p_1^2+p_2^2+p_3^3+2p_1p_2\cos\beta-2p_1p_2\cos\beta-2p_1p_3\cos\theta\\ -& 2p_2p_3(\sin\beta\sin\theta\cos\phi+\cos\beta\cos\theta))^{1/2}\,.
\end{align}
Moreover we write the $\delta$-distribution as
\begin{align}
    \delta(E_1+E_2-E_3-E_4)=\frac{E_1+E_2-E_3}{2p_2p_3\sin\beta\sin\theta}\delta(\cos\phi-\cos\phi_c)\,,
\end{align}
where
\begin{align}
    \cos\phi_c = \frac{m_1^2+m_2^2+m_3^2-m_4^2+2(E_1E_2-E_1E_3-E_2E_3)-2p_1p_2cos\beta}{2p_2p_3\sin\beta\sin\theta}
\end{align}
The integral over $\phi$ is symmetric around $\phi=\pi$, so we replace $\int_0^{2\pi}d\phi = 2\int_0^\pi d\phi$.
When the dust settles, we arrive at our reduced collision integral
\begin{align}
    \mathcal{C}_{(F)} =& 4(2\pi)^4 \int_0^\infty \frac{dp_1 p_1^2}{(2\pi)^3 2 E_1}
     \int_0^\infty \frac{dp_2 p_2^2}{(2\pi)^3 2 E_2} \int_0^\infty \frac{dp_3 p_3^2}{(2\pi)^3 2 E_3} \int_{-1}^1 d\cos\beta \int_{-1}^{1}d\cos\theta \\
    &\times \frac{1}{2E_4} \frac{E_1+E_2-E_3}{2p_2p_3\sin\beta\sin\theta} \times \Theta(0\leqslant\cos^2\phi_c \leqslant 1) \times F(\vec p_1,\vec p_2,\vec p_3,\vec p_4),,
    \label{eq:reduced_coll_22}
\end{align}
where we completed all the previously discussed steps and also evaluated the $\phi$ integration, i.e. $E_4$ and $\vec p_4$ are to be understood as functions of the remaining integration variables, and in particular $\cos\phi = \cos\phi_c$ from the transformed $\delta$-distribution is enforced. We restrict the integration domain to the physically allowed region by means of an indicator function $\Theta$. In principle, we could explicitly find the boundaries of the physically allowed integration domain (see e.g. \cite{Bringmann:2022aim}), but for our purposes this is not necessary.

The integrations $d\cos\beta$ and $d\cos\theta$ are already suitable for a Monte Carlo integration. For the spatial momentum, we make variable transformations $p_i=-\Lambda_i \log x_i$, where $x_i$ becomes our new integration variable and $\Lambda_i$ is a dimensional constant. We find good convergence behavior for a variety of choices, but typically use $\Lambda_i=\mathcal{O}(1-10)\,T_\SM$ for fast convergence (see also Reference \cite{Luo:2020fdt}). With these variable substitutions, Equation \ref{eq:reduced_coll_22} is the collision integral as we have implemented. We explicitly verified the correctness of the implementation against the well-known semi-analytic result from Gelmini and Gondolo \cite{Gondolo:1990dk}.

\subsection{$1\rightarrow2$ processes}
The integral is $9$ dimensional, and can be reduced to $2$, which would be perfectly suitable for traditional numerical integration. To illustrate the convenience of the Monte Carlo approach, we also perform the integration of (inverse) decays with Monte Carlo sampling. The steps are essentially the same as before, just with the added convenience of fewer integrals to handle.

We are once more free to choose the coordinate system. We make the same choice as before and use $\vec p_1 = p_1(0,0,1)^T$. As before, $\vec p_1$ and $\vec p_2$ form a plane, so it is possible to write $\vec p_2 = p_2(\sin\theta,0,\cos\theta)^T$ with trivial azimuthal dependence, i.e. we can pick the alignement of the plane to our convenience. The spatial part of the $\delta$-distribution eliminates $d^3p_3$, and now $E_3 = \sqrt{m_3^2+(\vec p_1-\vec p_2)^2}$ holds. For the remaining $\delta$-distribution, we write it in terms of the angle $\delta(\cos\theta-\cos\theta_c)$, where now
\begin{align}
    \cos\theta_c = \frac{m_3^2+p_1^2+p_2^2-(E_1-E_2)^2}{2p_1p_2},,
\end{align}
holds.
We restrict the integration domain to the physically allowed space by using an indicator function $\Theta(0\leqslant\cos^2\phi\leqslant1)$, but could equally well just derive an analytic expression for the integration domain. Putting everything together, we arrive at
\begin{align}
    \mathcal{C}_{(F)} = 8\pi^2 \int_0^\infty\frac{dp_1 p_1^2}{(2\pi)^32E_1} \int_0^\infty\frac{dp_2 p_2^2}{(2\pi)^32E_2} \times \frac{1}{2E_3} \frac{E_3}{p_1p_2} \times F(p_i)\,.
\end{align}
The remaining two integrations are again performed numerically after making the substitution $p_i=-\Lambda_i \log x_i$. We have also verified this collision integral against analytic decay rates when distributions are Maxwell-Boltzmann (see also Reference \cite{Luo:2020fdt}).

\subsection{Performance and other implementation details}
By first performing few adaptive integration steps of just below $1000$ evaluations, we can adapt the \texttt{VEGAS} integration routine to the integrand. The actual evaluation uses few integration steps but a larger number of evaluations. Using the internal uncertainty estimate of \texttt{VEGAS} and after optimizing some \texttt{VEGAS} configurations, we can achieve a relative error of $0.1\,\%$ for evaluation times on a single core in the $10-100\mathrm{\,ms}$ range (Precise values depend also on the underlying particle models). With suitable parallelization, an effective evaluation of many integrals, e.g. to provide an interpolation grid, remains feasible at manageable computational load.

Some processes can usually be added and jointly integrated, thus drastically reducing the number of integral evaluations even in the presence of many processes that contribute. In the case study here, we may define for the process $f\bar f\rightarrow\nu_R\bar\nu_R$ as
\begin{equation}
    \text{Process}(f) \propto \abs{\mathcal{M}_{f\bar f\rightarrow \nu_R\bar\nu_R}}^2 \left(f_1 f_2 \bar f_3 \bar f_4 - f_3 f_4 \bar f_1 \bar f_2\right)\,
\end{equation}
where we omitted all the factors coming from variable transformations in Equation \ref{eq:reduced_coll_22}. The integration is just over angles and dimensionless $3$-momenta, meaning  \text{Process}(f) with all its omitted prefactors contains the full information about a given channel. Then, it proves convenient to use Equation \ref{eq:reduced_coll_22} not for a single $\text{Process}(f)$, but rather $\sum_f\text{Process}(f)$, leading to a single evaluation of Equation \ref{eq:reduced_coll_22} at the cost of a moderately more expensive integrand.

We would also like to note that in many situations parameters can be factored out, e.g. powers of couplings for many processes. Then, the collision operator only needs to be precomputed once, which again reduces the number of evaluations needed.

\section{Matrix elements, cross sections and decay rates}
\label{sec:matrix}
Here we list all relevant matrix elements. Expressions involving $\gamma$-matrices have been evaluated using \texttt{FeynCalc} \cite{Mertig:1990an,Shtabovenko:2020gxv}.

For the (inverse) decays we have
\begin{equation}
    \abs{\mathcal{M}_{\zprime \rightarrow f\bar f}}^2 = 4\left(2m_f^2 + m_\zprime^2\right)\,.
\end{equation}

\begin{equation}
    \abs{\mathcal{M}_{\zprime \rightarrow \nu_R\bar\nu_R}}^2 = 2m_\zprime^2\,.
\end{equation}
The total decay width in vaccum is given by
\begin{equation}
    \Gamma_\zprime = \frac{g^2}{12\pi}m_\zprime\left( 3 +  \sum_{f\neq\nu}^{m_\zprime>2m_f} \frac{{Q_{B-L}^{2}}}{N_c(f)} \left(1+\frac{2m_f}{m_\zprime}\right) \sqrt{1-\frac{4m_f^2}{m_\zprime^2}} \right)\,,
\end{equation}
where the factor $3$ comprises the contribution of left- and right-handed neutrinos. We neglect the complications of hadronization and assume decay into free quarks only.

For the process $f\bar f \leftrightarrow \nu_R \bar\nu_R$, we find
\begin{equation}
    \abs{\mathcal{M}_{f\bar f\rightarrow\nu_R\bar\nu_R}}^2 = g_\zprime^4  N_C(f)Q_f^2 Q_{\nu_R}^2 \frac{4 \left(2 m_f^4+2 m_f^2 (s-t-u)+t^2+u^2\right)}{\left(m_{Z^\prime}^2-s\right)^2}\,.
\end{equation}
The prescription for the resonant case is discussed in the main text. In the following, we show explicitly that in vacuum, the NWA and the explicit matrix element give the same cross section for resonant production.

We write the differential cross section as
\begin{equation}
    \frac{d\sigma}{dt} = \frac{1}{64\pi s} \frac{1}{\abs{\vec p_{1,\text{cm}}}^2} \abs{\mathcal{M}}^2\,,
\end{equation}
replace $u=2m_f^2-t-s$, and integrate the Mandelstam $t$ with boundaries $-(p_{1,\text{cm}}\mp p_{3,\text{cm}})$, where $p_{1,\text{cm}}=\sqrt{s/4-m_f^2}$ and $p_{3,\text{cm}}=\sqrt{s}/2$.
Then, after also multipliying by $3$  for the generations of $\nu_R$ in the final state, we find
\begin{equation}
    \sigma=\frac{g^4}{2\pi}\frac{Q_f^2 Q_{\nu_R}^2}{N_C(f)} \sqrt{\frac{s}{s-4m_f^2}} \left(1+\frac{2m_f^2}{s}\right) \frac{s}{(s-m_\zprime^2)^2+\Gamma_\zprime^2 m_\zprime^2}\,.
\end{equation}
Now if we approximate the resonance as a delta distribution, we readily arrive at
\begin{equation}
    \sigma_{\text{res}} = \frac{g^4}{2\pi}\frac{Q_f^2 Q_{\nu_R}^2}{N_C(f)} \sqrt{\frac{m_\zprime^2}{m_\zprime^2-4m_f^2}} \left(1+\frac{2m_f^2}{m_\zprime^2}\right) \frac{\pi m_\zprime}{\Gamma_\zprime} \delta(s-m_\zprime^2)\,.
\end{equation}
This result is to be compared with the corss section we find from the NWA prescription to judge whether spin correlations are relevant for the total cross section. Using Equation \ref{eq:NWA_vector} (The factor $1/3$ here is canceled by the three generations of $\nu_R$), the $t$ integration is now trivial, and we find 
\begin{equation}
    \sigma_\text{NWA} =  \frac{1}{64\pi s} \frac{1}{\abs{\vec p_{1,\text{cm}}}^2} g_\zprime^2 \frac{Q_f^2}{N_C(f)} g_\zprime^2 Q_{\nu_R}^2\,8 m_\zprime^2 (2m_f^2+m_\zprime^2) \frac{\pi}{m_\zprime\Gamma_\zprime} \sqrt{s}\sqrt{s-4m_f^2} \delta(s-m_\zprime^2)\,,
\end{equation}
where the last two square root factors come from the integration. After replacing $s=m_\zprime^2$ and simplifying, it is evident that $\sigma_{\text{NWA}}=\sigma_{\text{res}}$ holds, i.e. the total cross sections of the appraoches are the same. Therefore, spin correlations do not affect the total cross section in vacuum, and since the distributions we assume throughout this study do not explicitly depend on spin, we expect this to hold also for interactions in medium.


\bibliographystyle{JHEP}
\bibliography{biblio.bib}

\end{document}